\documentclass[reprint,amsmath,amssymb,aps,prl,showpacs]{revtex4-1}

\usepackage{graphicx}
\usepackage{dcolumn}
\usepackage{bm}

 \usepackage{color}

\usepackage{amssymb}
\usepackage{graphicx}
\usepackage{amsbsy}

\newcommand\beq{\begin{equation}}
\newcommand\eeq{\end{equation}}
\newcommand\beqa{\begin{eqnarray}}
\newcommand\eeqa{\end{eqnarray}}
\newcommand{\nn}{\nonumber\\}

\newcommand{\rr}{\mathbf{r}}

\newcommand{\la}{\lambda}

\newcommand{\dd}{\text{d}}
\newcommand{\ex}{\text{ex}}

\newcommand{\hs}{\text{HS}}
\newcommand{\py}{\text{PY-}}

\newcommand{\mo}{M_1}
\newcommand{\mt}{M_2}
\newcommand{\mth}{M_3}
\newcommand{\mn}{M_n}

\begin{document}



\title{Chemical-Potential Route: A Hidden Percus--Yevick Equation of State for Hard Spheres}


\author{Andr\'es Santos}
\email{andres@unex.es}
\homepage{http://www.unex.es/eweb/fisteor/andres/}

\affiliation{Departamento de F\'{\i}sica, Universidad de
Extremadura, Badajoz, E-06071, Spain}

\date{\today}

\begin{abstract}
The chemical potential of a hard-sphere fluid can be expressed in terms of the contact value of the radial distribution function of a solute particle with a diameter varying from zero to that of the solvent particles. Exploiting the explicit knowledge of such a contact value within the Percus--Yevick (PY) theory, and using standard thermodynamic relations, a hitherto unknown PY equation of state,  $p/\rho k_BT=-(9/\eta)\ln(1-\eta)-(16-31\eta)/2(1-\eta)^2$,  is unveiled. This equation of state  turns out to be  better than the one obtained from the conventional virial route. Interpolations  between the chemical-potential and compressibility routes  are shown to be more accurate than the widely used Carnahan--Starling equation of state. The extension  to polydisperse hard-sphere systems is also presented.

\end{abstract}


\maketitle

\paragraph*{Motivation and discussion.---}As is well known, the hard-sphere (HS) model  is of great importance in condensed matter, colloids science, and liquid state theory from both academic and practical points of view \cite{HM06,L01,M08}.
The model has also attracted a lot of interest because it provides a nice example of the rare existence of nontrivial exact solutions of an integral-equation theory, namely the Percus--Yevick (PY) theory \cite{PY58} for odd dimensions \cite{W63,W64,T63,L64,FI81,L84,RHS04,RHS07,RS07,RS11,RS11b}.

As generally expected from an approximate theory, the radial distribution function (RDF) provided by the PY integral equation suffers from thermodynamic inconsistencies; i.e., the thermodynamic quantities derived from the same RDF via different routes are not necessarily mutually consistent. In particular, the PY solution for three-dimensional one-component HSs of diameter $\sigma$ yields the following expression for the compressibility factor $Z\equiv p/\rho k_BT$ (where $p$ is the pressure, $\rho$ is the number density, $k_B$ is Boltzmann's constant, and $T$ is the temperature) through the virial (or pressure) route \cite{W63,W64,T63}:
\beq
Z_{\py v}(\eta)=\frac{1+2\eta+3\eta^2}{(1-\eta)^2}.
\label{PYv}
\eeq
Here, $\eta=\frac{\pi}{6}\rho\sigma^3$ is the packing fraction and the subscript $v$ is used to emphasize that the result corresponds to the virial route. In contrast, the compressibility route yields
\beq
Z_{\py c}(\eta)=\frac{1+\eta+\eta^2}{(1-\eta)^3}.
\label{PYc}
\eeq
Equation \eqref{PYc} is also obtained from the scaled-particle theory (SPT) \cite{RFL59,MR75,HC04b}.
The celebrated and accurate Carnahan--Starling (CS) \cite{CS69} equation of state (EOS) is obtained as the simple interpolation
\beqa
Z_{\text{CS}}(\eta)&=&\frac{1}{3}Z_{\py v}(\eta)+\frac{2}{3}Z_{\py c}(\eta)\nn
&=&\frac{1+\eta+\eta^2-\eta^3}{(1-\eta)^3}.
\label{CS}
\eeqa

For general interaction potentials, the third conventional route to the EOS is the energy route \cite{HM06}. However, this route is useless in the case of HSs since the internal energy is just that of an ideal gas, and thus it is independent of density. On the other hand, starting from a square-shoulder interaction and then taking the limit of vanishing shoulder width, it has been proved that the resulting HS EOS coincides exactly with the one obtained through
the virial route, regardless of the approximation used \cite{S05,S06}. Therefore, the energy and virial routes to the EOS
can be considered as equivalent in the case of HS fluids.

\begin{table*}
\caption{First ten virial coefficients $b_n$ as obtained exactly and from several EOS related to the PY theory.
\label{tab}}
\begin{ruledtabular}
\begin{tabular} {cccccccc}
$n$ &Exact&$Z_{\py v}$&$Z_{\py c}$&$Z_{\py \mu}$&$Z_{\text{CS}}$&$Z_{\mu c,1}$&$Z_{\mu c,2}$\\
\hline
${2} $ &$$4$$&$4$&$4$&$4$&$4$&$4$&$4$\\
$ {3} $ & $10$&$10$&$10$&$10$&$10$&$10$&$10$\\
$ {4} $ & $18.36476\ldots$&$16$&$19$&$16.75$&$18$&$18.1$&$18.125$\\
$ {5} $ & $28.2245$&$22$&$31$&$23.8$&$28$&$28.12$&$28.2$\\
${6} $ & $39.815$&$28$&$46$&$31$&$40$&$40$&$40.166\ldots$\\
$ {7} $ &$53.34$&$34$&$64$&$38.285714\ldots$&$54$&$53.714285\ldots$&$54$\\
$ {8} $ & $68.54$&$40$&$85$&$45.625$&$70$&$69.25$&$69.6875$\\
$ {9} $ & $85.81$&$46$&$109$&$53$&$88$&$86.6$&$87.222\ldots$\\
$ {10} $ & $105.8$&$52$&$136$&$60.4$&$108$&$105.76$&$106.6$\\
\end{tabular}
\end{ruledtabular}
\end{table*}

Except perhaps in the context of the SPT \cite{RFL59,MR75,HC04b}, little attention has been paid to a fourth route to the EOS of HSs: the chemical-potential route  \cite{note_12_06_2}. In particular, to the best of the author's knowledge, the possibility of obtaining the EOS via this route by exploiting the exact solution of the PY equation for HS mixtures \cite{L64} seems to have been overlooked. The main aim of this Letter is to fill this gap and derive the results
\beq
\frac{\mu_{\text{PY}}^\ex(\eta)}{k_BT}=7\eta\frac{1+\eta/14}{(1-\eta)^2}-\ln(1-\eta),
\label{muex}
\eeq
\beq
Z_{\py \mu}(\eta)=-9\frac{\ln(1-\eta)}{\eta}-8\frac{1-\frac{31}{16}\eta}{(1-\eta)^2},
\label{Zmu}
\eeq
where $\mu^\ex$ is the excess chemical potential and the subscript  $\mu$ in Eq.\ \eqref{Zmu} denotes that the compressibility factor is  obtained from Eq.\ \eqref{muex}.
Equation \eqref{Zmu} differs from Eqs.\ \eqref{PYv} and \eqref{PYc} in that it includes a logarithmic term and thus it is not purely algebraic. Nevertheless, $Z_{\py \mu}(\eta)$ is analytic at $\eta=0$ and provides well-defined values for the (reduced) virial coefficients $b_n$ defined by
\beq
Z(\eta)=1+\sum_{n=2}^\infty b_n \eta^{n-1}.
\label{vir}
\eeq
Table \ref{tab} compares the first ten virial coefficients obtained from the three PY EOS, Eqs.\ \eqref{PYv}, \eqref{PYc}, and \eqref{Zmu}, with the exact analytical ($n=2$--$4$) and Monte Carlo ($n=5$--$10$) values \cite{LKM05,CM06}. The interpolated coefficients obtained from Eq.\ \eqref{CS} are also included.
We observe that the virial coefficients $b_n^{(\py \mu)}$ obtained from Eq.\ \eqref{Zmu} are in general noninteger rational numbers. More explicitly, $b_n^{(\py\mu)}=(18-31n+15n^2)/2n$, while $b_n^{(\py v)}=2(3n-4)$, $b_n^{(\py c)}=(3n^2-3n+2)/2$, and   $b_n^{(\text{CS})}=(n+2)(n-1)$.

Interestingly enough, the virial coefficients from the chemical-potential route are more accurate than those from the virial route, although less than the ones from the compressibility route. This suggests the possibility of exploring  CS-like interpolations of the form $Z_{\mu c}(\eta)=\alpha Z_{\py \mu}(\eta)+(1-\alpha)Z_{\py c}(\eta)$ with $\alpha\approx 0.4$. Two simple and convenient choices are $\alpha=\frac{2}{5}$ and $\alpha=\frac{7}{18}$. Thus,
\beqa
Z_{\mu c,1}(\eta)&=&\frac{2}{5} Z_{\py \mu}(\eta)+\frac{3}{5}Z_{\py c}(\eta)\nn
&=&-\frac{18}{5}\frac{\ln(1-\eta)}{\eta}-\frac{13-50\eta+28\eta^2}{5(1-\eta)^3},
\label{muc1}
\eeqa
\beqa
Z_{\mu c,2}(\eta)&=&\frac{7}{18} Z_{\py \mu}(\eta)+\frac{11}{18}Z_{\py c}(\eta)\nn
&=&-\frac{7}{2}\frac{\ln(1-\eta)}{\eta}-\frac{30-117\eta+65\eta^2}{12(1-\eta)^3}.
\label{muc2}
\eeqa
The values $b_n^{(\mu c,1)}=(36-56n+21n^2+9n^3)/10n$ and $b_n^{(\mu c,2)}=(42-65n+24n^2+11n^3)/12n$  obtained from Eqs.\ \eqref{muc1} and \eqref{muc2}, respectively, are also displayed in Table \ref{tab}. We observe a very good agreement, even better than that of $b_n^{(\text{CS)}}$, with the exact values, especially in the case of $b_n^{(\mu c,1)}$. In particular, $b_{10}^{(\mu c,1)}$ is excellent.

\begin{figure}
  \includegraphics[width=.9\columnwidth]{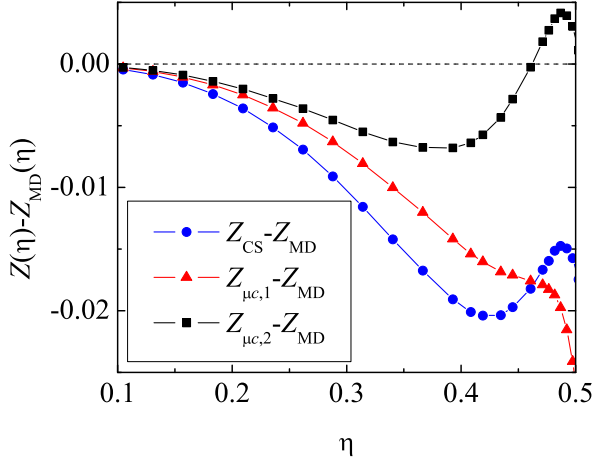}
\caption{(color online). Plot of $Z_{\text{CS}}(\eta)-Z_{\text{MD}}(\eta)$ (circles), $Z_{\mu c, 1}(\eta)-Z_{\text{MD}}(\eta)$ (triangles), and  $Z_{\mu c, 2}(\eta)-Z_{\text{MD}}(\eta)$ (squares).} \label{fig}
\end{figure}

The superiority of $Z_{\mu c,1}$ and $Z_{\mu c,2}$ over $Z_{\text{CS}}$ is confirmed by Fig.\ \ref{fig}, where the differences $Z_{\text{CS}}(\eta)-Z_{\text{MD}}(\eta)$, $Z_{\mu c, 1}(\eta)-Z_{\text{MD}}(\eta)$, and  $Z_{\mu c, 2}(\eta)-Z_{\text{MD}}(\eta)$ (where $Z_{\text{MD}}$ denotes molecular dynamics simulation values \cite{KLM04}) are compared. As can be seen, both $Z_{\mu c,1}$ and $Z_{\mu c,2}$ deviate from $Z_{\text{MD}}$ less than $Z_{\text{CS}}$   over most of the stable liquid region. It is noteworthy that, while $Z_{\mu c,1}$ predicts better virial coefficients than $Z_{\mu c,2}$, the latter EOS is more accurate for $\eta>0.1$.

In the case of  a \emph{polydisperse} HS fluid characterized by a size distribution $x(\sigma')$, it will be  proved elsewhere \cite{note_12_06} that Eqs.\ \eqref{muex} and \eqref{Zmu} are generalized to
\beqa
\frac{\mu_\text{PY}^\ex(\eta,\sigma)}{k_BT}&=&-\ln(1-\eta)+\frac{\eta}{1-\eta}\left[
\frac{\sigma^3}{\mth}
+3\frac{\mo\sigma^2}{\mth}\right.\nn
&& \left(1+\frac{\eta}{1-\eta}\frac{\mt\sigma}{\mth}\right)
+3\frac{\mt\sigma}{\mth}\nn
&&\left.\times\left(1+\frac{3}{2}\frac{\eta}{1-\eta}\frac{\mt\sigma}{\mth}\right)\right],
\label{r2}
\eeqa
\beqa
Z_{\py \mu}(\eta)&=&\frac{1}{1-\eta}+\frac{\mo\mt}{\mth}\frac{3\eta}{(1-\eta)^2}-9\frac{\mt^3}{\mth^2}\nn
&&\times\left[\frac{\ln(1-\eta)}{\eta}+
\frac{1-\frac{3}{2}\eta}{(1-\eta)^2}\right],
\label{r3}
\eeqa
where $\mn\equiv\int_0^\infty\dd \sigma'\,x(\sigma')\sigma'^n$ is the $n$th moment of the size distribution and in Eq.\ \eqref{r2} $\mu_\text{PY}^\ex(\eta,\sigma)$ is the excess chemical potential of spheres of diameter $\sigma$. Equation \eqref{r3} differs from $Z_{\py v}$ and $Z_{\py c}$ for mixtures \cite{L64} only by the coefficient of $\mt^3/\mth^2$. Analogously to the one-component case, it is possible to construct interpolated EOS $Z_{\mu c}=\alpha Z_{\py\mu}+(1-\alpha)Z_{\py c}$, with $\alpha=\frac{2}{5}$ or $\alpha=\frac{7}{18}$,  which are more accurate than the Boubl\'ik--Mansoori--Carnahan--Starling--Leland EOS \cite{B70,MCSL71}.

Once the main features and applications of the new PY EOS \eqref{Zmu} have been discussed, the rest of the Letter is devoted to its derivation.

\paragraph*{The chemical-potential route.---}
Let us consider a $d$-dimensional system made of $N$ hard spheres ($i=1,\ldots,N$) of diameter $\sigma$ (the ``solvent'') plus one ``solute'' particle ($i=0$) which interacts with the solvent particles via a HS potential of core $\lambda\sigma$ with $0\leq\lambda\leq 1$. Thus, the total potential energy function is
\beqa
\Phi_{N+1}(\rr^{N+1};\lambda)&\equiv&\Phi_{N+1}(\rr_0,\rr_1,\ldots,\rr_N;\lambda)\nn
&=&\sum_{j=1}^N\phi_\hs(r_{0j}/\lambda)+\sum_{1\leq i<j\leq N}\phi_\hs(r_{ij}),\nn
\label{1}
\eeqa
where $r_{ij}=|\rr_i-\rr_j|$ and $\phi_\hs(r)=\infty$ if $r<\sigma$ and $0$ otherwise. Thus,
\beq
e^{-\beta\Phi_{N+1}(\rr^{N+1};\lambda)}=\prod_{j=1}^N \Theta(r_{0j}-\lambda\sigma)\prod_{1\leq i<j\leq N}\Theta(r_{ij}-\sigma),
\label{2}
\eeq
where $\beta\equiv 1/k_BT$ and $\Theta(x)$ is  the Heaviside step function. We will further need the property
\beq
\frac{\partial}{\partial\lambda}e^{-\beta\Phi_{N+1}(\rr^{N+1};\lambda)}=-\sigma e^{-\beta\Phi_{N+1}(\rr^{N+1};\lambda)}\sum_{j=1}^N\delta(r_{0j}-\lambda\sigma^+).
\label{6}
\eeq

The configuration integral of the solvent+solute system is defined by
\beqa
Q_{N+1}(V;\lambda)&=&V^{-(N+1)}\int \dd\mathbf{r}_0\int \dd\mathbf{r}_1\ldots\int \dd\mathbf{r}_N\,\nn
&&\times e^{-\beta\Phi_{N+1}(\rr^{N+1};\lambda)},
\label{QN}
\eeqa
where $V$ is the volume. Note that $Q_{N+1}(V;0)=Q_N(V)$ is  the configuration integral of the $N$-particle system of solvent particles. Likewise, $Q_{N+1}(V;1)=Q_{N+1}(V)$ is the configuration integral of a normal system of $N+1$ identical particles. Therefore, one can write \cite{RFL59}
\beqa
\frac{\mu^\ex}{k_BT}&=&-\ln\frac{Q_{N+1}(V)}{Q_{N}(V)}\nn
&=&-\int_0^1\dd\lambda\,\frac{\partial}{\partial\lambda}\ln Q_{N+1}(V;\lambda).
\label{11}
\eeqa

If $\lambda\leq \frac{1}{2}$, the solute-solvent HS interaction is \emph{nonadditive} since the solute can ``penetrate'' the hard core of radius $\frac{1}{2}\sigma$. Provided   the solvent particles are in a nonoverlapping configuration, the volume they exclude to the position of the solute particle is simply $N v_d(2\lambda\sigma)^d$, i.e.,
\beq
\int\dd\rr_0\, \prod_{j=1}^N \Theta(r_{0j}-\lambda\sigma)=V-N v_d(2\lambda\sigma)^d,
\label{12}
\eeq
where $v_{d}=(\pi /4)^{d/2}/\Gamma (1+d/2)$ is the volume of a
$d$-dimensional sphere of unit diameter.
Consequently,
\beq
Q_{N+1}(V;\lambda)=\left[1-(2\lambda)^d\eta\right]Q_N(V),\quad \lambda\leq \frac{1}{2},
\label{13}
\eeq
where $\eta=\rho v_d\sigma^d$ is the packing fraction of the solvent system, $\rho=N/V$ being the number density. Equation \eqref{13} allows one to rewrite Eq.\ \eqref{11} as
\beqa
\frac{\mu^\ex}{k_BT}&=&-\ln\frac{Q_{N+1}(V;\frac{1}{2})}{Q_{N}(V)}-\ln\frac{Q_{N+1}(V)}{Q_{N+1}(V;\frac{1}{2})}\nn
&=&-\ln(1-\eta)-\int_{\frac{1}{2}}^1\dd\lambda\,\frac{\partial}{\partial\lambda}\ln Q_{N+1}(V;\lambda).
\label{14}
\eeqa

Next, we take into account that the RDF for the solute particle is defined as
\beqa
g(r_{01};\lambda,\eta)&=&
\frac{V^{-(N-1)}}{Q_{N+1}(V;\lambda)} \int \dd\mathbf{r}_2\ldots\int \dd\mathbf{r}_N\nn
&&\times e^{-\beta\Phi_{N+1}(\rr^{N+1};\lambda)}.
\label{3}
\eeqa
Application of Eq.\ \eqref{6} into Eq.\ \eqref{QN} then yields
\beqa
\frac{\partial}{\partial\lambda}\ln Q_{N+1}(V;\lambda)
&=&-\sigma\rho\int\dd\rr\,\delta(r-\lambda\sigma^+)g(r;\lambda,\eta)\nn
&=&-d 2^d\lambda^{d-1}\eta g(\lambda;\eta),
\label{9}
\eeqa
where in the second equality   $g(\lambda;\eta)\equiv g(r=\lambda\sigma^+;\lambda,\eta)$.
{}From Eqs.\ \eqref{13} and \eqref{9} we obtain the exact result
\beq
g(\lambda;\eta)=\frac{1}{1-(2\lambda)^d\eta},\quad \lambda\leq \frac{1}{2}.
\label{15}
\eeq

Let us now consider Eq.\ \eqref{14}. Inserting Eq.\ \eqref{9},
\beq
\frac{\mu^\ex(\eta)}{k_BT}=-\ln(1-\eta)+d 2^d\eta\int_{\frac{1}{2}}^1\dd\lambda\,\lambda^{d-1}g(\lambda;\eta).
\label{16}
\eeq
The thermodynamic relation
\beq
\left(\frac{\partial p}{\partial\rho}\right)_T=\rho\left(\frac{\partial\mu}{\partial\rho}\right)_T
\label{therm}
\eeq
yields
\beq
\frac{\partial}{\partial\eta}\left[\eta Z_\mu(\eta)\right]=\frac{1}{1-\eta}+d2^d\eta\frac{\partial}{\partial\eta}
\left[\eta\int_{\frac{1}{2}}^1\dd\lambda\,\lambda^{d-1}g(\lambda;\eta)\right].
\label{18}
\eeq
Integrating both sides over density, we finally obtain
\beqa
Z_\mu(\eta)&=&-\frac{\ln(1-\eta)}{\eta}+d2^d\eta\int_{\frac{1}{2}}^1\dd\lambda\,\lambda^{d-1}g(\lambda;\eta)\nn
&&-
\frac{d2^d}{\eta}\int_0^\eta\dd\eta'\eta'\int_{\frac{1}{2}}^1\dd\lambda\,\lambda^{d-1}g(\lambda;\eta').
\label{19}
\eeqa
This constitutes the chemical-potential route to the EOS of the HS fluid. As in the virial route
\beq
Z_v(\eta)=1+2^{d-1}\eta g(1;\eta),
\label{17}
\eeq
Eq.\ \eqref{19} requires the \emph{contact} value of the RDF, not the full spatial dependence (as required by the compressibility route). On the other hand, in contrast to Eq.\ \eqref{17}, Eq.\ \eqref{19} is ``nonlocal'' in the sense that it needs the knowledge of the contact value of a \emph{solute} particle of diameter $\sigma_0=(2\lambda-1)\sigma$ in the range $0\leq \sigma_0\leq \sigma$ and, moreover, for packing fractions $\eta'<\eta$.

Equation \eqref{19} is formally exact. We now specialize to the three-dimensional case ($d=3$) and consider the PY approximation for $g(\lambda;\eta)$, namely \cite{L64}
\beq
g_{\text{PY}}(\lambda;\eta)=\frac{1}{1-\eta}+\frac{3}{2}\frac{\eta}{(1-\eta)^2}\frac{2\la-1}{\la},\quad \lambda\geq \frac{1}{2}.
\label{25}
\eeq
Insertion of the above expression into the right-hand sides of Eqs.\ \eqref{16}, \eqref{19}, and \eqref{17} finally provides Eqs.\ \eqref{muex}, \eqref{Zmu}, and \eqref{PYv}, respectively.

The fact that Eq.\ \eqref{Zmu} is more accurate than Eq.\ \eqref{PYv} can be explained by the following argument. Equations   \eqref{16}, \eqref{18}, and \eqref{19} show that the chemical-potential route is directly related to the integral
\beq
\int_{\frac{1}{2}}^1\dd\lambda\,\lambda^{d-1}g(\lambda;\eta),
\label{average}
\eeq
which is proportional to a  (weighted) \emph{average} of the contact value $g(\lambda;\eta)$ in the  range $\frac{1}{2}\leq \lambda\leq 1$. Since both the contact value $g(\lambda;\eta)$ and its first derivative $\partial g(\lambda;\eta)/\partial \lambda$ at $\lambda=\frac{1}{2}$ are given exactly by the PY equation [compare Eqs.\ \eqref{15} and \eqref{25}], it seems reasonable that the ``average'' value \eqref{average} is better estimated than the end point at $\la=1$ by the PY approximation.

Thus far, all the results have been specialized to HS systems. In the more general case of particles interacting through a potential $\phi(r)$, one can still single out a particle $i=0$ which interacts with the rest via a potential $\phi(r;\lambda)$ such that $\phi(r;0)=0$ and $\phi(r;1)=\phi(r)$. Proceeding in a similar way as before, one arrives at \cite{note_12_06}
\beq
{\mu^\ex}=d 2^dv_d\rho\int_{0}^1\dd\lambda\int_0^\infty\dd r\,r^{d-1}g(r;\lambda)\frac{\partial \phi(r;\lambda)}{\partial \lambda} .
\label{r1}
\eeq
In this equation the $\lambda$-protocol $\phi(r;\lambda)$ remains arbitrary. If $\phi(r)$ is not a singular potential, an obvious choice is  $\phi(r;\lambda)=\lambda\phi(r)$ \cite{R80}. However, this possibility is ill-defined if, as happens with $\phi_\hs(r)$, the potential diverges over a finite range. In that case, an adequate choice is  $\phi(r;\lambda)=\phi(r/\lambda)$. Using the identity $\partial_\lambda \phi(r;\lambda)=-k_BT e^{\beta \phi(r;\lambda)}\partial_\lambda e^{-\beta \phi(r;\lambda)}$ in Eq.\ \eqref{r1}, and particularizing to $\phi_\hs(r)$, the choice  $\phi(r;\lambda)=\phi(r/\lambda)$ yields Eq.\ \eqref{16}.

\paragraph*{Conclusion.---}
In summary, a hitherto hidden EOS for a HS fluid described by the PY liquid state theory, Eq.\ \eqref{Zmu}, has been unveiled. This new  EOS from the chemical-potential route competes favorably with the conventional one from the virial route by the reasons outlined above. Thus, at least in the framework of the PY theory, the chemical-potential route should be placed on the same footing as the standard virial and compressibility routes. Apart from its intrinsic academic and pedagogical interest, the new EOS has a practical impact. For instance, the chemical-potential and compressibility routes  allow for the construction of  interpolation proposals, Eqs.\ \eqref{muc1} and \eqref{muc2}, which are more accurate than the widely used CS EOS.
Moreover, the use of Eq.\ \eqref{16} for HS fluids and Eq.\ \eqref{r1} for more general systems can be very helpful for the construction of accurate EOS.
Extensions of this work to sticky hard spheres \cite{B68} and to hyperspheres \cite{FI81,L84,RS07} are planned.

I am indebted to D. J. Henderson, J. Kolafa, E. Lomba, M. L\'opez de Haro, L. L. Lee, and A. Malijevsk\'y for helpful comments.  Financial support from the Spanish Government through Grant No. FIS2010-16587 and from the Junta de Extremadura (Spain) through Grant No.\ GR10158 (partially financed by FEDER funds) is  acknowledged.

\bibliographystyle{apsrev}
\bibliography{D:/Dropbox/Public/bib_files/liquid}
\begin{thebibliography}{30}
\bibitem[{not({\natexlab{a}})}]{note_12_06_2}
\bibinfo{note}{This should not be confused with the chemical potential obtained
  from zero-separation theorems. See, for instance, P. V. Giaquinta, G. Giunta,
  and G. Malescio, J. Stat. Phys. \textbf{63}, 141 (1991).}
\end{thebibliography}
\end{document}